\documentclass[conference]{IEEEtran}
\IEEEoverridecommandlockouts
\usepackage{cite}
\usepackage{amsmath,amssymb,amsfonts}
\usepackage{algorithmic}
\usepackage{graphicx}
\usepackage{textcomp}
\usepackage{xcolor}
\def\BibTeX{{\rm B\kern-.05em{\sc i\kern-.025em b}\kern-.08em
    T\kern-.1667em\lower.7ex\hbox{E}\kern-.125emX}}
\usepackage{balance}
\usepackage{algorithm,algorithmic}
\usepackage{float}
\usepackage[margin=2cm]{geometry}
\usepackage{amssymb,amsfonts}
\usepackage{amsmath}
\usepackage{amsthm}
\theoremstyle{definition}

\newtheorem{remark}{Remark}
\usepackage{color}
\usepackage{hyperref}

\begin{document}

\title{
Hybrid Vector Message Passing for Generalized Bilinear Factorization
}


\author{\IEEEauthorblockN{Hao Jiang\IEEEauthorrefmark{1}, Xiaojun Yuan\IEEEauthorrefmark{1}, and Qinghua Guo\IEEEauthorrefmark{2}}
\IEEEauthorblockA{\IEEEauthorrefmark{1}{National Key Lab. of Wireless Commun., Univ. of Elec. Sci. and Tech. of China, Chengdu, China}}
\IEEEauthorblockA{\IEEEauthorrefmark{2}{
School of ECTE, Univ. of Wollongong, Wollongong, NSW 2522, Australia}\\
Email: jh@std.uestc.edu.cn,
  xjyuan@uestc.edu.cn,
  qguo@uow.edu.au}
  \vspace{-0.7cm}
}

\maketitle

\begin{abstract}
In this paper, we propose a new message passing algorithm that utilizes hybrid vector message passing (HVMP) to solve the generalized bilinear factorization (GBF) problem. The proposed GBF-HVMP algorithm integrates expectation propagation (EP) and variational message passing (VMP) via variational free energy minimization, yielding tractable Gaussian messages. 
Furthermore, GBF-HVMP enables vector/matrix variables rather than scalar ones in message passing, resulting in a loop-free Bayesian network that improves convergence. Numerical results show that GBF-HVMP significantly outperforms state-of-the-art methods in terms of NMSE performance and computational complexity.

\end{abstract}


\section{Introduction}\label{intro}

A plethora of problems of interest in
science, engineering, and e-commerce can be formulated as 
estimating matrices $\boldsymbol{S} \in \mathbb{C}^{L \times K}$ and $\boldsymbol{X} \in \mathbb{C}^{K \times T}$ from certain measurements of their product $\boldsymbol{W} = \boldsymbol{S}\boldsymbol{X} \in \mathbb{C}^{L \times T}$. In many applications, the measurements are noisy and compressed/incomplete. These problems can be generally described by the canonical generalized bilinear factorization (GBF) model, i.e., to recover $\boldsymbol{S}$ and $\boldsymbol{X}$ from measurements given by
\begin{align}\label{model}
    \boldsymbol{y} = \mathcal{A}(\boldsymbol{S}\boldsymbol{X}) + \boldsymbol{n} \in \mathbb{C}^{N},
\end{align}
where $\mathcal{A}(\cdot):\mathbb{C}^{L \times T} \rightarrow \mathbb{C}^N$ is a linear operator, and $\boldsymbol{n} \in \mathbb{C}^N$ denotes an additive white Gaussian noise (AWGN) vector. The entries of $\boldsymbol{n}$ are i.i.d. drawn from $\mathcal{CN}(0,\sigma^2)$ with $\sigma^2$ being the noise power. The linear operator can be expressed in a matrix form as
$\mathcal{A}(\boldsymbol{S}\boldsymbol{X}) = \boldsymbol{A} \operatorname{vec}(\boldsymbol{SX})$,
where $\boldsymbol{A} \in \mathbb{C}^{N \times LT}$ is the matrix representation of $\mathcal{A}(\cdot)$, and $\operatorname{vec}(\cdot)$ denotes the vectorization operation. This model can be applied to a wide range of problems, depending on the nature of the linear operator $\mathcal{A}(\cdot)$. 
For example, in the blind channel-and-signal estimation problem \cite{zhang_blind_2018}, the linear operator $\mathcal{A}(\cdot)$ consists of steering vectors.
Another example is the compressive robust principle component analysis problem \cite{xue_turbo-type_2018}, where $\mathcal{A}(\cdot)$ is a compressive operator.

Greedy methods \cite{aharon2006k,waters_sparcs_2011}, optimization with convex relaxation \cite{wright_compressive_2013, aravkin_variational_2014}, and Bayesian inference based message passing algorithms \cite{liu2019super,6898015,xue_turbo-type_2018} are among the existing approaches to solve bilinear factorization problems. 
Message passing algorithms exhibit outstanding performance because they can better exploit the \textit{a priori} information of the factor variables.
However, there are two issues with the existing message passing algorithms. First, the existing message passing algorithms for bilinear factorization \cite{liu2019super,6898015} are derived from loopy belief propagation (LBP) \cite{heskes2002stable} with scalar variables, resulting in loopy Bayesian networks that deteriorate factorization performance. Second, the messages based on LBP are computationally intractable, leading to approximate message calculations with Taylor series expansion. These approximations may cause the message passing algorithms to diverge, so adaptive damping is used in these algorithms to improve convergence. Yet, adaptive damping slows down the convergence speed, and as a result, significantly increases the computational complexity.

To address the above difficulties, we develop a new message passing algorithm to tackle the GBF problem defined in \eqref{model}. Specifically, we first establish a probabilistic model for the GBF problem. Based on that, we derive a new message passing rule, named hybrid vector message passing (HVMP), for the GBF problem by following the principle of variational free energy minimization. HVMP effectively overcomes the challenges of the existing LBP based algorithms for bilinear factorization. HVMP employs vector/matrix variables in the derivation, resulting in a loop-free Bayesian network. Moreover, messages based on HVMP are tractable Gaussian distributions, thereby eliminating the need for 
Taylor series expansion. 
The linear operator $\mathcal{A}(\cdot)$ in the GBF problem may lead to operations with high computational complexity, such as the Kronecker product operation. By introducing marginalization to reduce the computational complexity, we propose the so-called GBF-HVMP algorithm based on the message passing rule of HVMP.
Numerical results demonstrate that
the GBF-HVMP algorithm, compared to the state-of-the-art, exhibits much better performance and faster convergence speed without needing adaptive damping.

\section{Message Passing Rule for HVMP}\label{MPrules}
We assume that the joint probability distribution of $\boldsymbol{y}$, $\boldsymbol{S}$ and $\boldsymbol{X}$ in \eqref{model} can be factorized as
\begin{equation}\label{probmodel0}
    \begin{aligned}
        p(\boldsymbol{y}, \boldsymbol{S}, \boldsymbol{X}) = f_{\boldsymbol{y}}(\boldsymbol{y}, \boldsymbol{S}, \boldsymbol{X})f_{\boldsymbol{S}}(\boldsymbol{S})f_{\boldsymbol{X}}(\boldsymbol{X}),
    \end{aligned}
\end{equation}
where factor function $f_{\boldsymbol{y}}(\boldsymbol{y}, \boldsymbol{S}, \boldsymbol{X}) \triangleq p(\boldsymbol{y} | \boldsymbol{S},\boldsymbol{X})=\mathcal{CN}(\boldsymbol{y}; \boldsymbol{A}\operatorname{vec}(\boldsymbol{S} \boldsymbol{X}),\sigma^2\mathbf{I}_N)$ is the conditional pdf of $\boldsymbol{y}$ given $\boldsymbol{S}$ and $\boldsymbol{X}$, $f_{\boldsymbol{S}}(\boldsymbol{S}) \triangleq p(\boldsymbol{S})=\prod_{l,k}^{L,K}p(s_{l,k})$ and $f_{\boldsymbol{X}}(\boldsymbol{X}) \triangleq p(\boldsymbol{X})=\prod_{k,t}^{K,T}p(x_{k,t})$ are the \textit{a priori} distributions of $\boldsymbol{S}$ and $\boldsymbol{X}$, respectively. 

We next establish HVMP for the GBF problem via variational free energy minimization \cite{1459044}.
To start with, we introduce auxiliary distribution functions
$q_{f_{\boldsymbol{y}}}(\boldsymbol{y}, \boldsymbol{S}, \boldsymbol{X})$, $q_{f_{\boldsymbol{S}}}(\boldsymbol{S})$, and $q_{f_{\boldsymbol{X}}}(\boldsymbol{X})$ to approximate the statistical relations of variables $\boldsymbol{S}$ and $\boldsymbol{X}$ at the corresponding factor functions $f_{\boldsymbol{y}}(\boldsymbol{y}, \boldsymbol{S}, \boldsymbol{X})$, $f_{\boldsymbol{S}}(\boldsymbol{S})$, and $f_{\boldsymbol{X}}(\boldsymbol{X})$. We denote the auxiliary marginal distributions of matrix variables $\boldsymbol{S}$ and $\boldsymbol{X}$ by $q_{\boldsymbol{S}}(\boldsymbol{S})$ and $q_{\boldsymbol{X}}(\boldsymbol{X})$, respectively. Then, the Bethe approximation \cite{cowell1998advanced} of the joint probability distribution \eqref{probmodel0} can be expressed as
\begin{equation}\label{approxq}
    \begin{aligned}
        q(\boldsymbol{y}, \boldsymbol{S}, \boldsymbol{X}) = \frac{q_{f_{\boldsymbol{y}}}(\boldsymbol{y}, \boldsymbol{S}, \boldsymbol{X})q_{f_{\boldsymbol{S}}}(\boldsymbol{S})q_{f_{\boldsymbol{X}}}(\boldsymbol{X})}{q_{\boldsymbol{S}}(\boldsymbol{S})q_{\boldsymbol{X}}(\boldsymbol{X})}.
    \end{aligned}
\end{equation}
With the joint pdf in \eqref{probmodel0} and the Bethe approximation in \eqref{approxq}, we define the following variational free energy:
\begin{equation}\label{Fq}
    \begin{aligned}
        F(q) ={}& \int_{ \boldsymbol{S}, \boldsymbol{X}}q(\boldsymbol{y}, \boldsymbol{S}, \boldsymbol{X})\ln\frac{q(\boldsymbol{y}, \boldsymbol{S}, \boldsymbol{X})}{p(\boldsymbol{y}, \boldsymbol{S}, \boldsymbol{X})}\\ &\hspace{-4.5em}= \sum_{f\in\mathcal{F}}\int_{\mathcal{V}_f} q_{f}(\mathcal{V}_f) \ln \frac{q_{f}(\mathcal{V}_f)}{f(\mathcal{V}_f)} -\sum_{\boldsymbol{v}\in\mathcal{V}}\int_{\boldsymbol{v}} q_{\boldsymbol{v}}(\boldsymbol{v})\ln q_{\boldsymbol{v}}(\boldsymbol{v}),
    \end{aligned}
\end{equation}
where vector variable $\boldsymbol{v}\in\mathcal{V}\triangleq\{\operatorname{vec}(\boldsymbol{S}),\operatorname{vec}(\boldsymbol{X})\}$, factor $f\in\mathcal{F}\triangleq\{f_{\boldsymbol{y}}, f_{\boldsymbol{S}}, f_{\boldsymbol{X}}\}$, $\mathcal{V}_{f}$ represents the set of all variables associated with factor $f$, $f(\mathcal{V}_f)$ and $q_f(\mathcal{V}_f)$ are the factor function and the auxiliary distribution corresponding to factor $f$, respectively.
We next describe the constraints of the auxiliary distributions to be satisfied. We assume that $q_{f_{\boldsymbol{y}}}(\boldsymbol{y}, \boldsymbol{S}, \boldsymbol{X})$ can be factorized as the product of $q_{f_{\boldsymbol{y},\boldsymbol{S}}}(\boldsymbol{S})$ and $q_{f_{\boldsymbol{y},\boldsymbol{X}}}(\boldsymbol{X})$, i.e.,
\begin{equation}\label{partialqfz}
    \begin{aligned}
        q_{f_{\boldsymbol{y}}}(\boldsymbol{y},\boldsymbol{S},\boldsymbol{X}) = q_{f_{\boldsymbol{y},\boldsymbol{S}}}(\boldsymbol{S})q_{f_{\boldsymbol{y},\boldsymbol{X}}}(\boldsymbol{X}).
    \end{aligned}
\end{equation}
This additional factorization plays a pivotal role in distinguishing HVMP from other existing message passing rules, as elaborated later in Remark \ref{remark1}. The Bethe approximation requires the auxiliary distributions to fulfill the marginalization consistency constraint \cite{1459044}. 
For the sake of tractability, we relax the marginalization consistency constraint into first- and second-order moment matching of the auxiliary distributions \cite{minka2001ep2}, i.e., for any $f\in\mathcal{F}$, $\boldsymbol{v}\in\mathcal{V}_f$,
\begin{subequations}\label{rmcc}
    \begin{align}
        \operatorname{E}\left[\boldsymbol{v}\bigg|\int_{\mathcal{V}_f\setminus\boldsymbol{v}}q_f(\mathcal{V}_f)\right] ={}& \operatorname{E}\left[\boldsymbol{v}|q_{\boldsymbol{v}}(\boldsymbol{v})\right],\label{E-E2}\\
        \operatorname{E}\left[\boldsymbol{v}\boldsymbol{v}^\mathrm{H}\bigg|\int_{\mathcal{V}_f\setminus\boldsymbol{v}}q_f(\mathcal{V}_f)\right] ={}& \operatorname{E}\left[\boldsymbol{v}\boldsymbol{v}^\mathrm{H}|q_{\boldsymbol{v}}(\boldsymbol{v})\right],\label{Var-Var2}
    \end{align}
\end{subequations}
where $\operatorname{E}[\cdot|q(\cdot)]$ denotes the expectation over the distribution $q(\cdot)$.
In addition, the auxiliary distributions satisfy the normalization constraint:
\begin{equation}\label{nc}
    \begin{aligned}
        \int q_{\iota}(\boldsymbol{v}) = 1, \quad \forall\iota\in \{f_{\boldsymbol{y}},f_{\boldsymbol{S}}, f_{\boldsymbol{X}}, \boldsymbol{S}, \boldsymbol{X} \}.
    \end{aligned}
\end{equation}
We thus formulate the optimization problem of minimizing the variational free energy as
\begin{equation}
\begin{aligned}\label{optimization}
    \mathop{\min}\limits_{q(\cdot)} \eqref{Fq} \quad \text{s.t.} \quad \eqref{partialqfz}-\eqref{nc}.
\end{aligned}
\end{equation}
Following \cite{1459044}, we can solve the optimization problem, with the message passing rule of HVMP summarized below:
\begin{subequations}\label{messages}
    \begin{align}
        m_{f \rightarrow \boldsymbol{v}}(\boldsymbol{v}) &= \frac{1}{m_{\boldsymbol{v} \rightarrow f}(\boldsymbol{v})}\operatorname{proj}\left[ m_{\boldsymbol{v} \rightarrow f}(\boldsymbol{v})\right. \nonumber \\ &\hspace{-2.5em}\left.\times e^{\int_{\mathcal{V}_f\setminus\boldsymbol{v}} \prod_{\boldsymbol{v}^\prime\in\mathcal{V}_f\setminus \boldsymbol{v}} m_{f \rightarrow \boldsymbol{v}^\prime}(\boldsymbol{v}^\prime) m_{\boldsymbol{v}^\prime \rightarrow f}(\boldsymbol{v}^\prime) \ln f(\mathcal{V}_f)}\right],\label{mfatoxii0}\\m_{\boldsymbol{v} \rightarrow f}(\boldsymbol{v}) &= \prod_{f^\prime\in\mathcal{F}_{\boldsymbol{v}} \setminus f} m_{f^\prime \rightarrow \boldsymbol{v}}(\boldsymbol{v}),\label{mxiitofa0}
    \end{align}
\end{subequations}
where $\mathcal{F}_{\boldsymbol{v}}$ is the set of all factors associated with variable $\boldsymbol{v}$, 
$\operatorname{proj}[p(\cdot)]$ denotes the projection of $p(\cdot)$ to a Gaussian distribution with matched first- and second-order moments. We omit the derivation of \eqref{messages} due to space limitation.

\begin{remark}\label{remark1}
HVMP has two major differences from the existing message passing algorithms. Firstly, HVMP employs vector/matrix variables rather than scalar ones, which accounts for the inclusion of ``vector'' in the name of HVMP, emphasizing the utilization of vector message passing. Secondly, the message passing rule outlined in \eqref{messages} can be interpreted as a hybrid of two conventional message passing algorithms, namely, expectation propagation (EP) \cite{minka2001ep2} and variational message passing (VMP) \cite{winn_variational_nodate}.
More specifically, we show how to relate HVMP with the conventional EP and VMP. Replace the additional factorization \eqref{partialqfz} with auxiliary marginal distributions of scalar variables, i.e.,
\begin{equation}\label{qxiiprod}
    \begin{aligned}
        q_{\boldsymbol{v}}(\boldsymbol{v}) = \prod_i q_{v_i}(v_i), \forall \boldsymbol{v}\in\mathcal{V},
    \end{aligned}
\end{equation}
where $v_{i}$ is the $i$-th scalar element of variable $\boldsymbol{v}$. We then 
minimize \eqref{Fq} subject to \eqref{rmcc}, \eqref{nc} and \eqref{qxiiprod}
by following the steps in \cite{1459044}. The resulting messages are
\begin{subequations}
    \begin{align}
        m_{f \rightarrow v_i}(v_i) ={}& \frac{1}{m_{v_i \rightarrow f}(v_i)}\operatorname{proj}\left[ \int_{\mathcal{V}_f\setminus v_i} f(\mathcal{V}_f) \right.\nonumber\\ &\left.\prod_{v_j\in\mathcal{V}_f\setminus v_i} m_{v_j \rightarrow f}(v_j)  \right],\label{mftoviEP}\\ m_{v_i \rightarrow f}(v_i) ={}& \prod_{f^\prime\in\mathcal{F}_{v_i}\setminus f}m_{f^\prime \rightarrow v_i}(v_i),\label{mvitofEP}
    \end{align}
\end{subequations}
which is exactly the message passing rule of EP \cite[eq. (83) and (54)]{minka2005divergence}. On the other hand, we may assume that all the auxiliary distributions are factorized into distributions of scalar variables, i.e.,
\begin{equation}\label{qfprod}
    \begin{aligned}
        q_{\iota}(\boldsymbol{v}) = \prod_iq_{\iota_i}(v_i), \forall\iota\in \{f_{\boldsymbol{y}}, f_{\boldsymbol{S}}, f_{\boldsymbol{X}}, \boldsymbol{S}, \boldsymbol{X} \}.
    \end{aligned}
\end{equation}
Minimizing \eqref{Fq} subject to \eqref{rmcc}, \eqref{nc} and \eqref{qfprod}, we obtain the message from factor $f$ to variable $v_i$ as
\begin{equation}\label{mftoviVMP}
    \begin{aligned}
        m_{f \rightarrow v_i}(v_i) &= e^{\int_{\mathcal{V}_f\setminus v_i} \prod_{v_j\in\mathcal{V}_f\setminus v_i} m_{f \rightarrow v_j}(v_j) m_{v_j \rightarrow f}(v_j) \ln f(\mathcal{V}_f)},
    \end{aligned}
\end{equation}
and the message from variable $v_i$ to factor $f$ as in \eqref{mvitofEP}. This gives the message passing rule of VMP \cite[eq. (71) and (54)] {minka2005divergence}. We observe that the HVMP rule \eqref{mfatoxii0} is a certain combined form of \eqref{mftoviEP} and \eqref{mftoviVMP}, i.e., HVMP is a hybrid of EP and VMP (with vector variables).
\end{remark}

\begin{figure}[htb]
    \centering
    \vspace{-0.4cm}
    \includegraphics[width=0.35\textwidth]{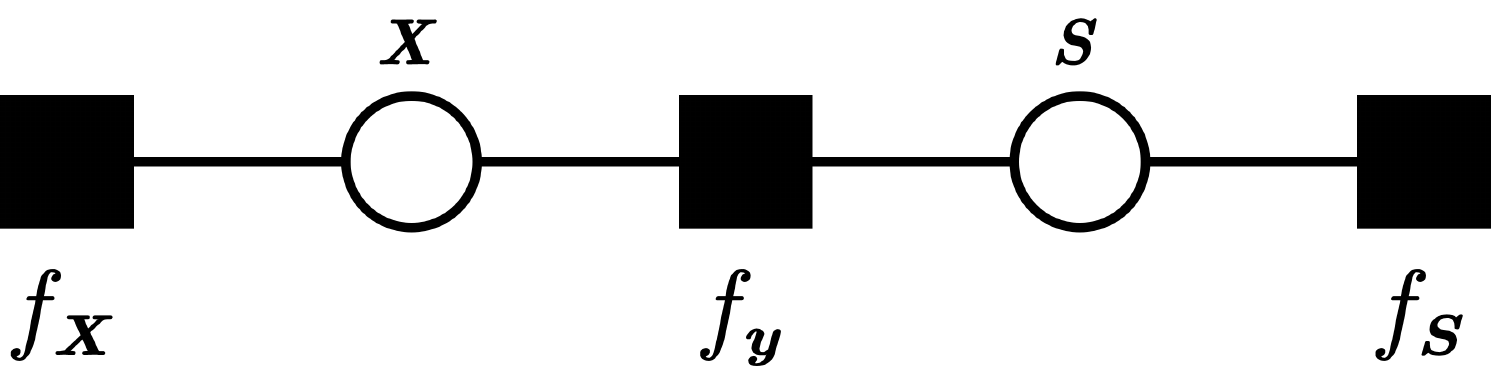}
    \vspace{-0.3cm}
    \caption{A factor graph representation of \eqref{probmodel0}, where the variable nodes represent vector/matrix variables rather than scalar ones. The corresponding message passing rule is defined by \eqref{messages}.}
    \label{factorgraph}
    \vspace{-0.4cm}
\end{figure}

\begin{remark}\label{remark2}
Many algorithms, such as the BiG-AMP algorithm \cite{6898015} and the P-BiG-AMP algorithm \cite{parker_parametric_2016}, have been developed to solve bilinear factorization problems via message passing. Notably, these algorithms are based on the principle of LBP \cite{heskes2002stable}
which is a special case of EP with some additional approximations.
However, these bilinear factorization algorithms suffer from the difficulties discussed in Section \ref{intro}.
HVMP effectively overcomes these difficulties:
\begin{itemize}
    \item By introducing vector/matrix variables into the derivation of HVMP, we achieve a loop-free Bayesian network for the GBF problem as illustrated in Fig. \ref{factorgraph}.
    \item With the additional factorization in \eqref{partialqfz}, the hybrid message passing rule of HVMP results in tractable Gaussian messages, as detailed in Section \ref{Algdesign}.
    \item Adaptive damping is not necessary for the GBF-HVMP algorithm to ensure convergence, which significantly improves the computational efficiency.
\end{itemize}

\end{remark}

\section{Algorithm Design for GBF-HVMP}\label{Algdesign}
\subsection{Factor Graph and Messages}
For a better understanding of HVMP, we present a factor graph representation of \eqref{probmodel0} shown in Fig. \ref{factorgraph}. 
In contrast to conventional factor graphs, the variable nodes in Fig. \ref{factorgraph} represent vector/matrix variables, and the messages along the edges are defined by \eqref{messages}. More specifically, 
messages between $\boldsymbol{X}$ and its neighboring factor nodes are
\begin{subequations}\label{messageX}
    \begin{align}
        m_{f_{\boldsymbol{y}} \rightarrow \boldsymbol{X}}(\boldsymbol{X}) ={}& \frac{1}{m_{\boldsymbol{X} \rightarrow f_{\boldsymbol{y}}}(\boldsymbol{X})}\operatorname{proj}\left[m_{\boldsymbol{X} \rightarrow f_{\boldsymbol{y}}}(\boldsymbol{X}) \right.\nonumber\\&\left.\times e^{\int_{\boldsymbol{S}} m_{\boldsymbol{S}}(\boldsymbol{S}) \ln f_{\boldsymbol{y}}(\boldsymbol{y},\boldsymbol{S},\boldsymbol{X})}\right],\label{mfztoX}\\m_{\boldsymbol{X} \rightarrow f_{\boldsymbol{X}}}(\boldsymbol{X}) ={}& m_{f_{\boldsymbol{y}} \rightarrow \boldsymbol{X}}(\boldsymbol{X}),\label{mxiitofa}\\m_{f_{\boldsymbol{X}} \rightarrow \boldsymbol{X}}(\boldsymbol{X}) ={}& \frac{1}{m_{\boldsymbol{X} \rightarrow f_{\boldsymbol{X}}}(\boldsymbol{X})}\operatorname{proj}\left[ m_{\boldsymbol{X} \rightarrow f_{\boldsymbol{X}}}(\boldsymbol{X}) f_{\boldsymbol{X}}({\boldsymbol{X}}) \right],\label{mfatoxii}\\m_{\boldsymbol{X} \rightarrow f_{\boldsymbol{y}}}(\boldsymbol{X}) ={}& m_{f_{\boldsymbol{X}} \rightarrow \boldsymbol{X}}(\boldsymbol{X}),\label{mxiitofz}\\
        m_{\boldsymbol{X}}(\boldsymbol{X}) ={}& m_{f_{\boldsymbol{y}} \rightarrow \boldsymbol{X}}(\boldsymbol{X})m_{\boldsymbol{X} \rightarrow f_{\boldsymbol{y}}}(\boldsymbol{X})\label{mv}
    \end{align}
\end{subequations} 
where \eqref{mv}
is the message of variable $\boldsymbol{X}$. Combining equations \eqref{mxiitofa}-\eqref{mv}, we obtain
\begin{equation}\label{mX}
    \begin{aligned}
        m_{\boldsymbol{X}}(\boldsymbol{X}) = \operatorname{proj}\left[m_{f_{\boldsymbol{y}} \rightarrow \boldsymbol{X}}(\boldsymbol{X}) f_{\boldsymbol{X}}({\boldsymbol{X}}) \right].
    \end{aligned}
\end{equation}
Similarly, messages between $\boldsymbol{S}$ and its neighboring factor nodes can be obtained by replacing $\boldsymbol{X}$ with $\boldsymbol{S}$ in \eqref{messageX}, i.e.,
    \begin{align}
        m_{f_{\boldsymbol{y}} \rightarrow \boldsymbol{S}}(\boldsymbol{S}) ={}& \frac{1}{m_{\boldsymbol{S} \rightarrow f_{\boldsymbol{y}}}(\boldsymbol{S})}\operatorname{proj}\left[m_{\boldsymbol{S} \rightarrow f_{\boldsymbol{y}}}(\boldsymbol{S}) \right.\nonumber\\&\left.\times e^{\int_{\boldsymbol{S}} m_{\boldsymbol{X}}(\boldsymbol{X}) \ln f_{\boldsymbol{y}}(\boldsymbol{y},\boldsymbol{S},\boldsymbol{X})}\right],\label{mfztoS}\\
        m_{\boldsymbol{S}}(\boldsymbol{S}) ={}& \operatorname{proj}\left[m_{f_{\boldsymbol{y}} \rightarrow \boldsymbol{S}}(\boldsymbol{S}) f_{\boldsymbol{S}}({\boldsymbol{S}}) \right].\label{mS}
    \end{align}
Hence, it suffices to only compute \eqref{mfztoX}, \eqref{mX}-\eqref{mS} in the message passing process. 
Since all the messages are Gaussian distributions, we present the explicit forms of the messages in Table \ref{table1} for clarity. 
Detailed explanations will be presented in the following subsections.
\begin{table}[htb]
    \centering
    \caption{Explicit Forms of Messages}
    \begin{tabular}{c|c}
        \hline Messages & Explicit form  \\ \hline
        $m_{f_{\boldsymbol{y}} \rightarrow \boldsymbol{X}}(\boldsymbol{X})$ & $\mathcal{CMN}(\boldsymbol{X};\overline{\boldsymbol{X}}, \overline{\boldsymbol{\Sigma}}_{\boldsymbol{X}}, \mathbf{I}_T)$\\ 
        $m_{\boldsymbol{X}}(\boldsymbol{X})$ & $\mathcal{CMN}(\boldsymbol{X};\hat{\boldsymbol{X}},\boldsymbol{U}_{\boldsymbol{X}},\mathbf{I}_T)$\\ 
        $m_{f_{\boldsymbol{y}} \rightarrow \boldsymbol{S}}(\boldsymbol{S})$ & $\mathcal{CMN}(\boldsymbol{S};\overline{\boldsymbol{S}},\mathbf{I}_L,\overline{\boldsymbol{\Sigma}}_{\boldsymbol{S}})$\\
        $m_{\boldsymbol{S}}(\boldsymbol{S})$ & $\mathcal{CMN}(\boldsymbol{S};\hat{\boldsymbol{S}},\mathbf{I}_L,\boldsymbol{V}_{\boldsymbol{S}})$\\ \hline
    \end{tabular}
    \label{table1}
\end{table}

\subsection{Computation of \texorpdfstring{$m_{f_{\boldsymbol{y}} \rightarrow \boldsymbol{X}}(\boldsymbol{X})$}{}}\label{sec-fz-fx}
From \eqref{mfztoX}, 
letting $\boldsymbol{x}=\operatorname{vec}(\boldsymbol{X})$, we obtain the following complex vector Gaussian form:
\begin{equation}\label{mfztoXpro}
    m_{f_{\boldsymbol{y}} \rightarrow \boldsymbol{X}}(\boldsymbol{X}) = \mathcal{CN}\left(\boldsymbol{x} ; \overline{\boldsymbol{x}},  \overline{\boldsymbol{\Sigma}}_x\right),
\end{equation}
with
\begin{subequations}\label{xbarandSigmax}
    \begin{align}
        \overline{\boldsymbol{x}}={}&\overline{\boldsymbol{\Sigma}}_x (\mathbf{I}_T \otimes \hat{\boldsymbol{S}})^{\mathrm{H}}\boldsymbol{A}^{\mathrm{H}} \boldsymbol{y}/\sigma^2,\label{xbar}\\
        \overline{\boldsymbol{\Sigma}}_x={}&\sigma^2\left((\mathbf{I}_T \otimes \hat{\boldsymbol{S}})^{\mathrm{H}} \boldsymbol{A}^{\mathrm{H}} \boldsymbol{A} (\mathbf{I}_T \otimes \hat{\boldsymbol{S}})\right.\nonumber\\&\left.+\boldsymbol{\mathring{A}}^\mathrm{H}\boldsymbol{\mathring{A}} \otimes \boldsymbol{V}_S\right)^{-1},\label{Sigmax}
    \end{align}
\end{subequations}
where $\boldsymbol{\mathring{A}}=\left[\operatorname{vec}(\boldsymbol{A}_{(1)}),\cdots,\operatorname{vec}(\boldsymbol{A}_{(T)})\right]$, $\boldsymbol{A}_{(i)}=\boldsymbol{A}(:,iL-L+1:iL))\in\mathbb{C}^{N\times L}$ and $\otimes$ denotes the Kronecker product. 
The detailed derivation of \eqref{mfztoXpro} can be found in Appendix \ref{ap-ztox}. Note that the Kronecker products and matrix inversion in \eqref{xbarandSigmax} require computational complexity $\mathcal{O}(NLK^2T)$. To alleviate the computational burden, we propose to approximately evaluate the message $m_{f_{\boldsymbol{y}} \rightarrow \boldsymbol{X}}(\boldsymbol{X})$ by using the following complex matrix Gaussian form:
\begin{equation}\label{mfztoX2}
    \begin{aligned}
    m_{f_{\boldsymbol{y}} \rightarrow \boldsymbol{X}}(\boldsymbol{X}) = \mathcal{CMN}\left(\boldsymbol{X} ; \overline{\boldsymbol{X}}, \overline{\boldsymbol{\Sigma}}_X, \mathbf{I}_T\right),
    \end{aligned}
\end{equation}
with
\begin{subequations}\label{matrixXandSigmaX}
    \begin{align}
        \overline{\boldsymbol{X}} &= \overline{\boldsymbol{\Sigma}}_X\hat{\boldsymbol{S}}^{\mathrm{H}} \hat{\boldsymbol{W}}/\nu_w,\label{WhatinX}\\
        \overline{\boldsymbol{\Sigma}}_X&=\nu_w\left(\hat{\boldsymbol{S}}^{\mathrm{H}}\hat{\boldsymbol{S}} + L\boldsymbol{V}_S\right)^{-1}.\label{SigmaX}
    \end{align}
\end{subequations}
In the above, $\hat{\boldsymbol{w}}=\operatorname{vec}(\hat{\boldsymbol{W}})$ and $\nu_w$ are given by
\begin{subequations}\label{lmmse}
    \begin{align}
        \hat{\boldsymbol{w}} &= \overline{\boldsymbol{w}}+\overline{\nu}_w\boldsymbol{A}^{\mathrm{H}}(\overline{\nu}_w\boldsymbol{A}\boldsymbol{A}^{\mathrm{H}}+\sigma^2\mathbf{I}_N)^{-1}(\boldsymbol{y}- \boldsymbol{A}\operatorname{vec}(\hat{\boldsymbol{S}}\hat{\boldsymbol{X}})),\label{lmmsewhat}\\
        \nu_w &= \frac{1}{LT}\operatorname{tr}\left(\overline{\nu}_w^2\boldsymbol{A}^{\mathrm{H}}(\overline{\nu}_w\boldsymbol{A}\boldsymbol{A}^{\mathrm{H}}+\sigma^2\mathbf{I}_N)^{-1}\boldsymbol{A}\right),\label{lmmsenuw}
    \end{align}
\end{subequations}
where $\overline{\boldsymbol{w}}$ and $\overline{\nu}_w$ are expressed as
\begin{subequations}\label{overlinew}
\begin{align}
    \overline{\boldsymbol{w}} &=\operatorname{vec}(\hat{\boldsymbol{S}}\hat{\boldsymbol{X}}),\\
    \overline{\nu}_w &= 2\sigma^2\frac{N}{\left\|\boldsymbol{A}\right\|_F^2}.
\end{align}
\end{subequations}
The detailed derivation of \eqref{mfztoX2} can be found in Appendix \ref{ap-approx}. 
The computational complexity of \eqref{matrixXandSigmaX}-\eqref{overlinew} is dominated by the matrix inversion in \eqref{lmmsewhat}, which requires $\mathcal{O}(N^3)$ multiplications. This is significantly lower than that of \eqref{xbarandSigmax}.

\subsection{Computation of \texorpdfstring{$m_{\boldsymbol{X}}(\boldsymbol{X})$}{}}\label{sec-mx}
From \eqref{mX}, 
the message of $\boldsymbol{X}$ is $m_{\boldsymbol{X}}(\boldsymbol{X}) = \operatorname{proj}\left[\mathcal{CMN}(\boldsymbol{X} ; \overline{\boldsymbol{X}}, \overline{\boldsymbol{\Sigma}}_X, \mathbf{I}_T)p(\boldsymbol{X})\right]$.
The mean and variance of 
$m_{\boldsymbol{X}}(\boldsymbol{X})$
are calculated as
\begin{subequations}\label{Xpost0}
    \begin{align}
        \hat{\boldsymbol{X}} ={}& \frac{1}{C_{X}}\int_{\boldsymbol{X}} \boldsymbol{X}\mathcal{CMN}(\boldsymbol{X} ; \overline{\boldsymbol{X}}, \overline{\boldsymbol{\Sigma}}_X, \mathbf{I}_T)p(\boldsymbol{X}),\label{xhat0}\\
        \nu_{x_{kt}} ={}& \frac{1}{C_{X}}\int_{\boldsymbol{X}} \left[\left|x_{kt}-\hat{x}_{kt}\right|^{2} p(\boldsymbol{X})\right.\nonumber\\&\left.\times\mathcal{CMN}(\boldsymbol{X} ; \overline{\boldsymbol{X}}, \overline{\boldsymbol{\Sigma}}_X, \mathbf{I}_T)\right],\forall k,t\label{nux0}
    \end{align}
\end{subequations}
where $C_{X} = \int_{\boldsymbol{X}} \mathcal{CMN}(\boldsymbol{X}; \overline{\boldsymbol{X}}, \overline{\boldsymbol{\Sigma}}_X, \mathbf{I}_T)p(\boldsymbol{X})$ and covariances between scalar variables are ignored for simplicity.

It is difficult to compute \eqref{Xpost0} due to the intractable integrals. 
To avoid this difficulty, we marginalize $m_{f_{\boldsymbol{y}} \rightarrow \boldsymbol{X}}(\boldsymbol{X})$ as follows.
Recall from \eqref{matrixXandSigmaX}, we can treat $\overline{\boldsymbol{X}}$ as an estimate of $\boldsymbol{X}$. We can model $\boldsymbol{X}$ as
\begin{equation}\label{obserX}
    \overline{\boldsymbol{X}} = \boldsymbol{X} + \Delta\boldsymbol{X},
\end{equation}
where $\Delta\boldsymbol{X}\sim\mathcal{CMN}\left(\Delta\boldsymbol{X};\boldsymbol{0}_{K,T}, \overline{\boldsymbol{\Sigma}}_X, \mathbf{I}_T\right)$ denotes the error of $\boldsymbol{X}$.
We employ the whitening transformation to \eqref{obserX} and obtain
\begin{equation}\label{AMP}
    {\overline{\boldsymbol{\Sigma}}_X}^{-\frac{1}{2}}\overline{\boldsymbol{X}} = {\overline{\boldsymbol{\Sigma}}_X}^{-\frac{1}{2}}\boldsymbol{X} + \boldsymbol{E}_X,
\end{equation}
where the entries of error $\boldsymbol{E}_X\in \mathbb{C}^{K\times T}$ are i.i.d. drawn from $\mathcal{CN}(0,1)$. From \eqref{AMP}, the element-wise mean and variance of $\boldsymbol{X}$ can be estimated by approximate message passing (AMP).
Distribution $\mathcal{CMN}(\boldsymbol{X} ; \overline{\boldsymbol{X}}, \overline{\boldsymbol{\Sigma}}_X, \mathbf{I}_T)$ is then approximated as
\begin{equation}\label{mfztoXfinal}
    \begin{aligned}
        \hspace{-6pt}\mathcal{CMN}(\boldsymbol{X} ; \overline{\boldsymbol{X}}, \overline{\boldsymbol{\Sigma}}_X, \mathbf{I}_T) \approx \prod_k^K\prod_t^T\mathcal{CN}\left(x_{kt} ; \underline{x}_{kt}, \underline{\nu}_{x_{kt}}\right), 
    \end{aligned}
\end{equation}
where $\underline{x}_{kt}$ and $\underline{\nu}_{x_{kt}}$ are obtained by AMP. Substituting \eqref{mfztoXfinal} into \eqref{Xpost0}, the mean and variance can be computed in an element-wise manner.
We average the variances of the elements in each row of $\boldsymbol{X}$, implying that each column of $\boldsymbol{X}$ shares the same covariance matrix
\begin{equation}\label{UX}
    \begin{aligned}
        \boldsymbol{U}_X=\frac{1}{T}\operatorname{Diag}\left(\left[\sum_t^T\nu_{x_{1t}},\cdots,\sum_t^T\nu_{x_{Kt}}\right]\right).
    \end{aligned}
\end{equation}
The message of $\boldsymbol{X}$ can thus be expressed as
\begin{equation}\label{mXpost}
    \begin{aligned}
        m_{\boldsymbol{X}}(\boldsymbol{X}) = \mathcal{CMN}(\boldsymbol{X};\hat{\boldsymbol{X}}, \boldsymbol{U}_X,\mathbf{I}_T).
    \end{aligned}
\end{equation}
The above whitening process can be skipped when the factor function $f_{\boldsymbol{X}}(\boldsymbol{X})=p(\boldsymbol{X})$ is a Gaussian distribution. In that case, we can directly combine $f_{\boldsymbol{X}}(\boldsymbol{X})$ with $m_{f_{\boldsymbol{y}} \rightarrow \boldsymbol{X}}(\boldsymbol{X})$,
as their product is still Gaussian.

\subsection{Computation of \texorpdfstring{$m_{f_{\boldsymbol{y}} \rightarrow \boldsymbol{S}}(\boldsymbol{S})$}{}}\label{sec-fz-fs}

From \eqref{mfztoS}, 
letting $\boldsymbol{s}=\operatorname{vec}(\boldsymbol{S})$, we obtain the following complex vector Gaussian form:
\begin{equation}\label{mfztoSpro}
    m_{f_{\boldsymbol{y}} \rightarrow \boldsymbol{S}}(\boldsymbol{S}) = \mathcal{CN}\left(\boldsymbol{s} ;\overline{\boldsymbol{s}}, \overline{\boldsymbol{\Sigma}}_s\right),
\end{equation}
with
\begin{subequations}\label{mfztoS2}
    \begin{align}
        \overline{\boldsymbol{s}}&=\overline{\boldsymbol{\Sigma}}_s (\hat{\boldsymbol{X}}^\top \otimes \mathbf{I}_L)^{\mathrm{H}} \boldsymbol{A}^{\mathrm{H}} \boldsymbol{y}/\sigma^2,\label{sbar}\\\overline{\boldsymbol{\Sigma}}_s={}&\sigma^2\left((\hat{\boldsymbol{X}}^\top \otimes \mathbf{I}_L)^{\mathrm{H}} \boldsymbol{A}^{\mathrm{H}} \boldsymbol{A} (\hat{\boldsymbol{X}}^\top \otimes \mathbf{I}_L)\right.\nonumber\\&\left.+\boldsymbol{U}_X \otimes \left(\sum_i^T\boldsymbol{A}_{(i)}^{\mathrm{H}} \boldsymbol{A}_{(i)}\right) \right)^{-1}.\label{Sigmasbar}
    \end{align}
\end{subequations}
Similarly to $m_{f_{\boldsymbol{y}} \rightarrow \boldsymbol{X}}(\boldsymbol{X})$, we reduce the computational complexity in \eqref{mfztoSpro}-\eqref{mfztoS2} by using the following complex matrix Gaussian form:
\begin{equation}\label{mfztoSmatrix}
    \begin{aligned}
    m_{f_{\boldsymbol{y}} \rightarrow \boldsymbol{S}}(\boldsymbol{S}) = \mathcal{CMN}\left(\boldsymbol{S} ; \overline{\boldsymbol{S}}, \mathbf{I}_L, \overline{\boldsymbol{\Sigma}}_S\right),
    \end{aligned}
\end{equation}
with
\begin{subequations}\label{SbarandnuS}
    \begin{align}
        \overline{\boldsymbol{S}} &= \hat{\boldsymbol{W}} \hat{\boldsymbol{X}}^{\mathrm{H}} \overline{\boldsymbol{\Sigma}}_S/\nu_w,\label{WhatinS}\\ \overline{\boldsymbol{\Sigma}}_S&=\nu_w\left(\hat{\boldsymbol{X}}\hat{\boldsymbol{X}}^{\mathrm{H}} + T\boldsymbol{U}_X\right)^{-1}.
    \end{align}
\end{subequations}

\subsection{Computation of \texorpdfstring{$m_{\boldsymbol{S}}(\boldsymbol{S})$}{}}\label{sec-ms}
From \eqref{mS}, 
the message of $\boldsymbol{S}$ is $m_{\boldsymbol{S}}(\boldsymbol{S}) =\operatorname{proj}\left[\mathcal{CMN}(\boldsymbol{S}; \overline{\boldsymbol{S}}, \mathbf{I}_L, \overline{\boldsymbol{\Sigma}}_S)p(\boldsymbol{S})\right]$.
The mean and variance of 
$m_{\boldsymbol{S}}(\boldsymbol{S})$
are calculated as
\begin{subequations}\label{Spost0}
    \begin{align}
        \hat{\boldsymbol{S}} ={}& \frac{1}{C_{S}}\int_{\boldsymbol{S}} \boldsymbol{S}\mathcal{CMN}(\boldsymbol{S}; \overline{\boldsymbol{S}}, \mathbf{I}_L, \overline{\boldsymbol{\Sigma}}_S)p(\boldsymbol{S}),\label{shat0}\\ 
        \nu_{s_{lk}} ={}& \frac{1}{C_{S}}\int_{\boldsymbol{S}} \left[\left|s_{lk}-\hat{s}_{lk}\right|^{2}p(\boldsymbol{S})\right.\nonumber \\&\left.\times\mathcal{CMN}(\boldsymbol{S}; \overline{\boldsymbol{S}}, \mathbf{I}_L, \overline{\boldsymbol{\Sigma}}_S)\right],\forall l,k\label{nus0}
    \end{align}
\end{subequations}
where $C_{S} = \int_{\boldsymbol{S}} \mathcal{CMN}(\boldsymbol{S}; \overline{\boldsymbol{S}}, \mathbf{I}_L, \overline{\boldsymbol{\Sigma}}_S)p(\boldsymbol{S})$.
Similarly to $m_{\boldsymbol{X}}(\boldsymbol{X})$, we marginalize $m_{f_{\boldsymbol{y}} \rightarrow \boldsymbol{S}}(\boldsymbol{S})$ as follows.
For message $\mathcal{CMN}(\boldsymbol{S}; \overline{\boldsymbol{S}}, \mathbf{I}_L, \overline{\boldsymbol{\Sigma}}_S)$, we model the mean $\overline{\boldsymbol{S}}$ as
\begin{equation}\label{obserS}
    \overline{\boldsymbol{S}} = \boldsymbol{S} + \Delta\boldsymbol{S},
\end{equation}
where $\Delta\boldsymbol{S}\sim\mathcal{CMN}(\Delta\boldsymbol{S}; \boldsymbol{0}_{L,K}, \mathbf{I}_L, \overline{\boldsymbol{\Sigma}}_S)$ denotes the error of $\boldsymbol{S}$.
By taking a similar whitening process as introduced in \ref{sec-mx}, we obtain
\begin{equation}\label{AMPS}
    {\overline{\boldsymbol{\Sigma}}_S}^{-\frac{1}{2}}\overline{\boldsymbol{S}}^{\mathrm{H}} = {\overline{\boldsymbol{\Sigma}}_S}^{-\frac{1}{2}}\boldsymbol{S}^{\mathrm{H}} + \boldsymbol{E}_S,
\end{equation}
where the entries of error $\boldsymbol{E}_S$ are i.i.d. drawn from $\mathcal{CN}(0,1)$. Estimating $\boldsymbol{S}$ from \eqref{AMPS} via AMP, $\mathcal{CMN}(\boldsymbol{S}; \overline{\boldsymbol{S}}, \mathbf{I}_L, \overline{\boldsymbol{\Sigma}}_S)$ is approximated and decoupled as
\begin{equation}\label{mfztoSfinal}
    \mathcal{CMN}(\boldsymbol{S}; \overline{\boldsymbol{S}}, \mathbf{I}_L, \overline{\boldsymbol{\Sigma}}_S) \approx \prod_l^L\prod_k^K\mathcal{CN}\left(s_{lk} ; \underline{s}_{lk}, \underline{\nu}_{s_{lk}}\right),
\end{equation}
where $\underline{s}_{lk}$ and $\underline{\nu}_{s_{lk}}$ are obtained by AMP. Substituting \eqref{mfztoSfinal} into \eqref{Spost0}, mean and variance can be computed in an element-wise manner.
We average the variances of the elements in each column of $\boldsymbol{S}$, meaning that the rows of $\boldsymbol{S}$ share the same covariance matrix
\begin{equation}\label{VS}
    \begin{aligned}
        \boldsymbol{V}_S=\frac{1}{L}\operatorname{Diag}\left(\left[\sum_l^L\nu_{s_{l1}},\cdots,\sum_l^L\nu_{s_{lK}}\right]\right).
    \end{aligned}
\end{equation}
The message of $\boldsymbol{S}$ can thus be expressed as
\begin{equation}\label{mSpost}
    \begin{aligned}
        m_{\boldsymbol{S}}(\boldsymbol{S}) = \mathcal{CMN}(\boldsymbol{S};\hat{\boldsymbol{S}}, \mathbf{I}_L, \boldsymbol{V}_S).
    \end{aligned}
\end{equation}

\subsection{Overall Algorithm}
The overall GBF-HVMP algorithm is summarized in Algorithm \ref{alg1}. We 
limit the max iteration number by $t_{\text{max}}$. $\hat{\boldsymbol{S}}$ and $\hat{\boldsymbol{X}}$ are initialized by drawing randomly from the \textit{a priori} distributions $p(\boldsymbol{S})$ and $p(\boldsymbol{X})$, respectively. 


\begin{algorithm}[htb]
    \caption{GBF-HVMP\label{alg1}}
    \begin{algorithmic}[0]
    \REQUIRE
    $\boldsymbol{y}$, $\sigma^2$, $\mathcal{A}(\cdot)$, $t_{\text{max}}$;
    \\
    \hspace{-10pt} 
    \textbf{Initialization:} 
    $\hat{\boldsymbol{S}}$, $\hat{\boldsymbol{X}}$, $\boldsymbol{V}_S$, $\boldsymbol{U}_X$;\\
    \hspace{-6pt}\textbf{for} {$t = 1,2,\cdots,t_{\text{max}}$} \textbf{do}
    \STATE \hspace{-0pt}1. Update auxiliary variables $\overline{\boldsymbol{w}}$, $\overline{\nu}_w$, $\hat{\boldsymbol{w}}$, $\nu_w$ in \eqref{lmmse}-\eqref{overlinew};
    \STATE \hspace{-0pt}2. Update $m_{f_{\boldsymbol{y}} \rightarrow \boldsymbol{X}}(\boldsymbol{X})$ by calculating $\overline{\boldsymbol{X}}$, $\overline{\boldsymbol{\Sigma}}_x$ in \eqref{matrixXandSigmaX};
    \STATE \hspace{-0pt}3. Update $m_{\boldsymbol{X}}(\boldsymbol{X})$ by calculating $\hat{\boldsymbol{X}}$, $\boldsymbol{U}_X$ in \eqref{Xpost0}-\eqref{UX};
    \STATE \hspace{-0pt}4. Update $m_{f_{\boldsymbol{y}} \rightarrow \boldsymbol{S}}(\boldsymbol{S})$ by calculating $\overline{\boldsymbol{S}}$, $\overline{\boldsymbol{\Sigma}}_s$ in \eqref{SbarandnuS};
    \STATE \hspace{-0pt}5. Update $m_{\boldsymbol{S}}(\boldsymbol{S})$ by calculating $\hat{\boldsymbol{S}}$, $\boldsymbol{V}_S$ in \eqref{Spost0}-\eqref{VS};
    \STATE \hspace{-6pt}\textbf{end for}
    \ENSURE $\hat{\boldsymbol{S}}$, $\hat{\boldsymbol{X}}$
    \end{algorithmic}
\end{algorithm}

We next discuss the computational complexity of the GBF-HVMP algorithm. 
The complexity of GBF-HVMP is dominated by the LMMSE estimation in step 1. This complexity can be reduced to $O(NLT)$ if the matrix form $\boldsymbol{A}$ of the linear operator $\mathcal{A}(\cdot)$ is pre-factorized using SVD 
or if $\boldsymbol{A}$ is partial-orthogonal, i.e., $\boldsymbol{AA}^\mathrm{H}=\mathbf{I}_N$. The complexity of the P-BiG-AMP algorithm \cite[Table \uppercase\expandafter{\romannumeral3}]{parker_parametric_2016} is $\mathcal{O}(NLTK^2)$ per iteration, which is much higher than that of GBF-HVMP. 
We emphasize that 
GBF-HVMP actually exhibits a much faster convergence speed than its counterpart message passing algorithms due to its advantages listed in Remark \ref{remark2}. 

\section{Numerical Results}\label{secWC}

We consider the case where $\boldsymbol{A}$ consists of randomly selected rows or columns of a discrete Fourier transformation matrix \cite{xue_turbo-type_2018}. $\boldsymbol{S}$ is a sparse matrix with the \textit{a priori} distribution being 
the product of Bernoulli-Gaussian distributions:
\begin{align}
    p(\boldsymbol{S})=\prod_{l}^{L}\prod_{k}^{K}\left((1-\rho)\delta(s_{l,k}) + \rho \mathcal{CN}(s_{l,k};0,1)\right),
\end{align}
where $\rho$ is the sparsity. Elements of $\boldsymbol{X}$ are i.i.d. generated from the standard complex Gaussian distribution, i.e.,
\begin{align}
    p(\boldsymbol{X})=\prod_{k}^{K}\prod_{t}^{T}\mathcal{CN}(x_{k,t};0,1).
\end{align}

\begin{figure}
    \centering
        \includegraphics[width=0.225\textwidth]{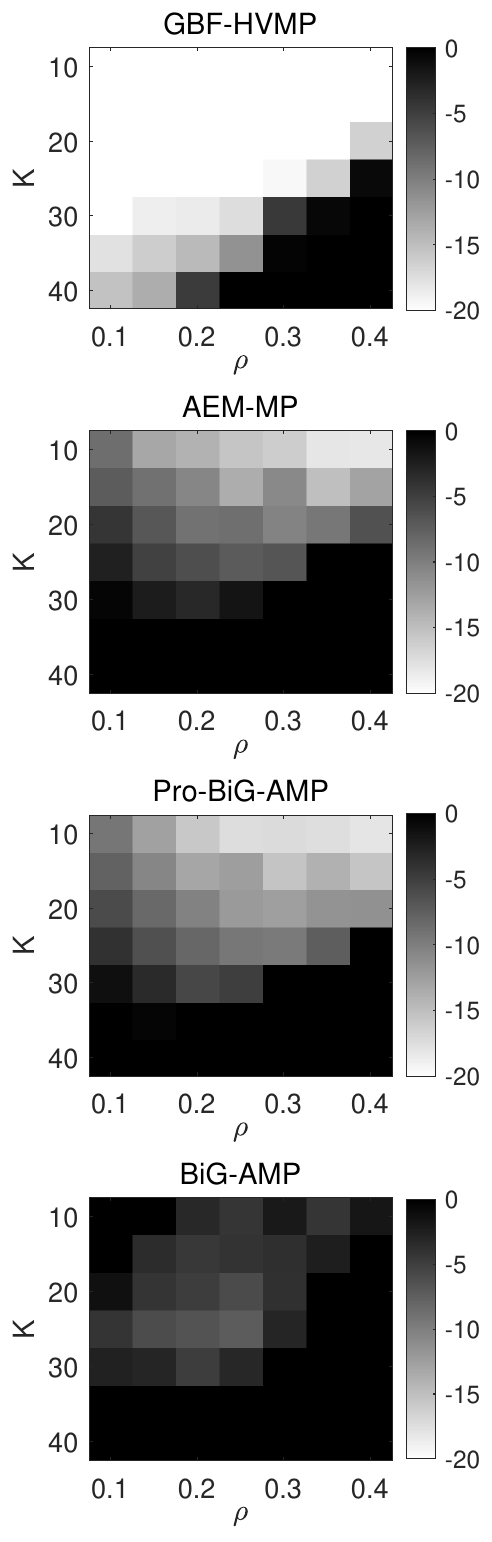}
        \includegraphics[width=0.225\textwidth]{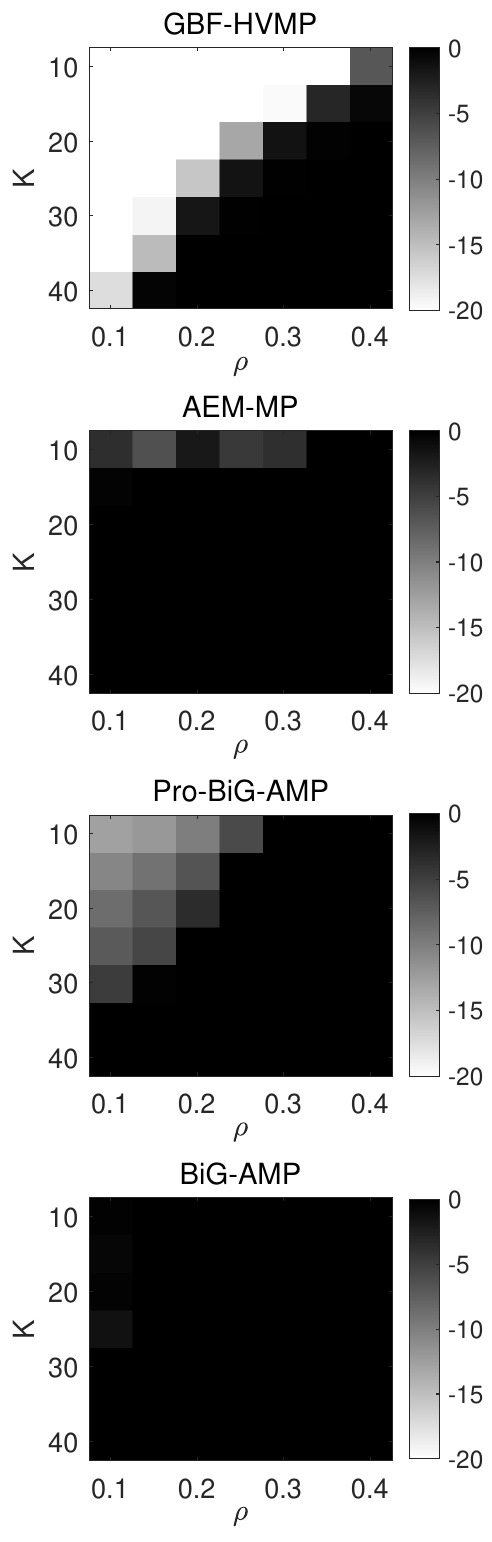}
    \caption{Phase transition of NMSE of $\boldsymbol{X}$, where $L=64$ for the four figures on the left and $L=256$ for the ones on the right. Other parameters are $N=128$, $T=50$ and SNR $=20\text{dB}$.}
    \label{PT-ALL}
    \vspace{-0.4cm}
\end{figure}

We compare the performance of GBF-HVMP with the baselines AEM-MP \cite{liu2019super}, BiG-AMP \cite{6898015} and Pro-BiG-AMP \cite{zhang_blind_2018}.
We show the normalized mean square error (NMSE) phase transitions of the algorithms in Fig. \ref{PT-ALL}. It depicts the NMSE of $\boldsymbol{X}$ with different sparsity $\rho$ and $K$. The SNR is fixed at $20$ dB. The left column corresponds to the case of $L=64$ and the right column corresponds to the case of $L=256$. In both cases, the GBF-HVMP algorithm exhibits a significantly better phase transition behavior. 
For example, if $L=64$ and $K=25$, GBF-HVMP can successfully recover $\boldsymbol{X}$ with NMSE $<-15$ dB when the sparsity $\rho<0.4$. 
But for the baselines, all the NMSEs fail to achieve $-15$ dB regardless of the sparsity.

\begin{figure}[htb]
    \centering
    \includegraphics[width=0.375\textwidth]{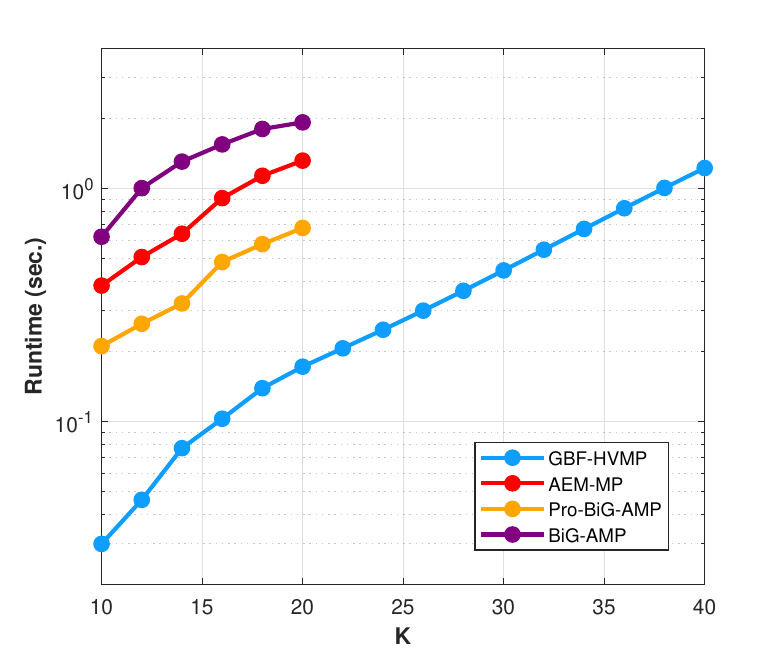}
    \caption{Running time versus $K$.
    Other parameters are $L=128$, $N = 128$, $T=50$, $\rho = 0.2$ and SNR $=20$ dB. Early stopping curves indicate that NMSEs of the corresponding algorithms cannot achieve $-20$ dB.}
    \label{runtime}
    \vspace{-0.3cm}
\end{figure}

We also evaluate the computational complexity of the algorithms in terms of running time. The SNR is fixed at $20$ dB. The stopping criterion is defined as the NMSE of $\boldsymbol{X}$ reaching $-20$ dB. 
As shown in Fig. \ref{runtime}, the curves of the baseline algorithms are stopped at $K=20$ because, for $K>20$, the NMSEs of the baselines cannot achieve $-20$ dB. GBF-HVMP achieves a much faster convergence speed compared to the baselines for any considered $K$.

\section{Conclusions}\label{conclusion}
In this paper, we proposed a new message passing algorithm named GBF-HVMP for the GBF problem based on the principle of variational free energy minimization. GBF-HVMP yields tractable Gaussian messages of vector/matrix variables, successfully addressing the difficulties in the existing LBP based message passing algorithms. Numerical results demonstrate that GBF-HVMP significantly outperforms the state-of-the-art methods in terms of both NMSE performance and computational complexity without needing adaptive damping.
\appendices
\section{Derivation of \texorpdfstring{\eqref{mfztoXpro}}{}}\label{ap-ztox}
Since $m_{\boldsymbol{S}}(\boldsymbol{S})$ is Gaussian distribution, the integral in \eqref{mfztoX} is calculated as
    \begin{align}
        &e^{\int_{\boldsymbol{S}} m_{\boldsymbol{S}}(\boldsymbol{S})\ln f_{\boldsymbol{y}}(\boldsymbol{y},\boldsymbol{S},\boldsymbol{X})} \label{intS2} \\
        \propto{}&e^{\left(-\frac{1}{\sigma^2} \int_{\boldsymbol{S}}\mathcal{CMN}(\boldsymbol{S};\hat{\boldsymbol{S}}, \mathbf{I}_L, \boldsymbol{V}_S)(\boldsymbol{s}^{\mathrm{H}} \tilde{\boldsymbol{X}}^{\mathrm{H}} \tilde{\boldsymbol{X}} \boldsymbol{s} - \boldsymbol{s}^{\mathrm{H}} \tilde{\boldsymbol{X}}^{\mathrm{H}} \boldsymbol{y}-\boldsymbol{y}^{\mathrm{H}} \tilde{\boldsymbol{X}} \boldsymbol{s})\right)},\nonumber
    \end{align}
where we use the identity  $\mathcal{A}\left(\boldsymbol{S} \boldsymbol{X}\right)=\boldsymbol{A}\left(\boldsymbol{X}^\top \otimes \mathbf{I}_L\right) \boldsymbol{s}$ and $\tilde{\boldsymbol{X}}=\boldsymbol{A}(\boldsymbol{X}^\top \otimes \mathbf{I}_L)$ in \eqref{intS2}. 
We next work out the integral in \eqref{intS2}. For the quadratic term $\boldsymbol{s}^{\mathrm{H}} \tilde{\boldsymbol{X}}^{\mathrm{H}} \tilde{\boldsymbol{X}} \boldsymbol{s}$, we have
\begin{equation}
    \begin{aligned}
        &\int_{\boldsymbol{S}} \mathcal{CMN}(\boldsymbol{S};\hat{\boldsymbol{S}}, \mathbf{I}_L, \boldsymbol{V}_S)\boldsymbol{s}^{\mathrm{H}} \tilde{\boldsymbol{X}}^{\mathrm{H}} \tilde{\boldsymbol{X}} \boldsymbol{s} \\
        ={}&\operatorname{tr}\left(\tilde{\boldsymbol{X}}^{\mathrm{H}} \tilde{\boldsymbol{X}}\left(\hat{\boldsymbol{s}} \hat{\boldsymbol{s}}^{\mathrm{H}}+\boldsymbol{V}_S \otimes \mathbf{I}_L\right)\right)\\
        ={}&\left\|\left(\left[\operatorname{vec}(\boldsymbol{A}_{(1)}),\cdots,\operatorname{vec}(\boldsymbol{A}_{(T)})\right] \otimes \boldsymbol{V}_S^{\frac{1}{2}}\right)\boldsymbol{x}\right\|_2^2\\&+\boldsymbol{x}^{\mathrm{H}}\left((\mathbf{I}_T \otimes \hat{\boldsymbol{S}})^{\mathrm{H}} \boldsymbol{A}^{\mathrm{H}} \boldsymbol{A}(\mathbf{I}_T \otimes \hat{\boldsymbol{S}})\right)\boldsymbol{x}\\
        ={}&\boldsymbol{x}^{\mathrm{H}}\left((\mathbf{I}_T \otimes \hat{\boldsymbol{S}})^{\mathrm{H}} \boldsymbol{A}^{\mathrm{H}} \boldsymbol{A}(\mathbf{I}_T \otimes \hat{\boldsymbol{S}}) + \boldsymbol{\mathring{A}}^\mathrm{H}\boldsymbol{\mathring{A}} \otimes \boldsymbol{V}_S\right)\boldsymbol{x},\label{intS5}
    \end{aligned}
\end{equation}
where $\boldsymbol{\mathring{A}}=\left[\operatorname{vec}(\boldsymbol{A}_{(1)}),\cdots,\operatorname{vec}(\boldsymbol{A}_{(T)})\right] $. 

Regarding terms $\boldsymbol{s}^{\mathrm{H}} \tilde{\boldsymbol{X}}^{\mathrm{H}} \boldsymbol{y}$ and $\boldsymbol{y}^{\mathrm{H}} \tilde{\boldsymbol{X}} \boldsymbol{s}$ in \eqref{intS2}, we have $\int_{\boldsymbol{S}} \mathcal{CMN}(\boldsymbol{S};\hat{\boldsymbol{S}}, \mathbf{I}_L, \boldsymbol{V}_S)\boldsymbol{s}^{\mathrm{H}} \tilde{\boldsymbol{X}}^{\mathrm{H}} \boldsymbol{y} = \boldsymbol{x}^{\mathrm{H}}\left(\mathbf{I}_T \otimes \hat{\boldsymbol{S}} \right)^{\mathrm{H}} \boldsymbol{A}^{\mathrm{H}} \boldsymbol{y}$, and $\int_{\boldsymbol{S}} \mathcal{CMN}(\boldsymbol{S};\hat{\boldsymbol{S}}, \mathbf{I}_L, \boldsymbol{V}_S)\boldsymbol{y}^{\mathrm{H}} \tilde{\boldsymbol{X}} \boldsymbol{s} =\boldsymbol{y}^{\mathrm{H}} \boldsymbol{A} \left(\mathbf{I}_T \otimes \hat{\boldsymbol{S}} \right) \boldsymbol{x}$. Based on the above results, \eqref{intS2} is proportional to a Gaussian distribution of $\boldsymbol{X}$. Noting that $m_{\boldsymbol{X} \rightarrow f_{\boldsymbol{y}}}(\boldsymbol{X})$ is a Gaussian distribution, we obtain
\begin{equation}
    \begin{aligned}
        m_{f_{\boldsymbol{y}} \rightarrow \boldsymbol{X}}(\boldsymbol{X})
        \propto {}&e^{\frac{1}{\sigma^2}\left( \boldsymbol{x}^{\mathrm{H}}\left(\mathbf{I}_T \otimes \hat{\boldsymbol{S}} \right)^{\mathrm{H}} \boldsymbol{A}^{\mathrm{H}} \boldsymbol{y} + \boldsymbol{y}^{\mathrm{H}} \boldsymbol{A} \left(\mathbf{I}_T \otimes \hat{\boldsymbol{S}} \right) \boldsymbol{x}\right)}\\&\hspace{-4em}\times e^{-\frac{1}{\sigma^2}\boldsymbol{x}^{\mathrm{H}}\left((\mathbf{I}_T \otimes \hat{\boldsymbol{S}})^{\mathrm{H}} \boldsymbol{A}^{\mathrm{H}} \boldsymbol{A}(\mathbf{I}_T \otimes \hat{\boldsymbol{S}}) + \boldsymbol{\mathring{A}}^{\mathrm{H}}\boldsymbol{\mathring{A}} \otimes \boldsymbol{V}_S\right)\boldsymbol{x}}.
    \end{aligned}
\end{equation}
This gives the Gaussian message in \eqref{mfztoXpro}.

\section{Derivation of \texorpdfstring{\eqref{matrixXandSigmaX}-\eqref{overlinew}}{}}\label{ap-approx}
To reduce computational complexity caused by 
the Kronecker products, we approximately evaluate $m_{f_{\boldsymbol{y}} \rightarrow \boldsymbol{X}}(\boldsymbol{X})$ via marginalization, where the correlations in each row of $\boldsymbol{X}$ are ignored.
Specifically,
recall that $f_{\boldsymbol{y}}(\boldsymbol{y},\boldsymbol{S},\boldsymbol{X})=\mathcal{CN}(\boldsymbol{y}; \boldsymbol{A}\operatorname{vec}(\boldsymbol{S} \boldsymbol{X}),\sigma^2\mathbf{I}_N)$, 
we have 
\begin{equation}\label{zzn}
    \begin{aligned}
        \boldsymbol{y} = \boldsymbol{A}\boldsymbol{w}+\boldsymbol{n},
        \vspace{-0.2cm}
    \end{aligned}
\end{equation}
where $\boldsymbol{n}\in\mathbb{C}^{N\times 1}$ with entries i.i.d. drawn form $\mathcal{CN}(0,\sigma^2)$. The random variable $\boldsymbol{w} = \operatorname{vec}(\boldsymbol{W})=\operatorname{vec}(\boldsymbol{SX})$ is with its mean and variance denoted by $\overline{\boldsymbol{w}}$ and $\overline{\nu}_w$. $\overline{\boldsymbol{w}}$ and $\overline{\nu}_w$ can be obtained from $m_{\boldsymbol{S}}(\boldsymbol{S})$ and $m_{\boldsymbol{X}}(\boldsymbol{X})$, with the explicit forms given in \eqref{overlinew}. From \eqref{zzn}, we have the linear minimum mean square error (LMMSE) estimation of $\boldsymbol{w}$ given in \eqref{lmmse}. The correlations between scalar variables in $\boldsymbol{w}$ are ignored via marginalization. With the LMMSE estimation $\hat{\boldsymbol{W}}$ and $\nu_w$, we have the model $\hat{\boldsymbol{W}} = \boldsymbol{SX} + \boldsymbol{N}_w$
where $\boldsymbol{N}_w\in\mathbb{C}^{L\times T}$ with entries i.i.d. drawn form $\mathcal{CN}(0,\nu_w)$. 
We thus obtain
\begin{align}
    &e^{\int_{\boldsymbol{S}} m_{\boldsymbol{S}}(\boldsymbol{S})\ln f_{\boldsymbol{y}}(\boldsymbol{y},\boldsymbol{S},\boldsymbol{X})}\nonumber\\ 
    \propto{}&e^{\left(- \int_{\boldsymbol{S}}\mathcal{CMN}(\boldsymbol{S};\hat{\boldsymbol{S}}, \mathbf{I}_L, \boldsymbol{V}_S)\operatorname{tr}\left(\left(\hat{\boldsymbol{W}}-\boldsymbol{SX}\right)^{\mathrm{H}}\left(\hat{\boldsymbol{W}}-\boldsymbol{SX}\right)/\nu_w\right)\right)}\nonumber \\
    \propto {}&e^{- \frac{1}{\nu_w}\operatorname{tr}\left(\boldsymbol{X}^{\mathrm{H}}\left(\hat{\boldsymbol{S}}^{\mathrm{H}}\hat{\boldsymbol{S}} + L \boldsymbol{V}_S\right)\boldsymbol{X} -\boldsymbol{X}^{\mathrm{H}}\hat{\boldsymbol{S}}^{\mathrm{H}}\hat{\boldsymbol{W}} - \hat{\boldsymbol{W}}^{\mathrm{H}}\hat{\boldsymbol{S}}\boldsymbol{X}\right)}.\label{intSS}
\end{align}
Substituting \eqref{intSS} into \eqref{mfztoX},
we obtain matrix-form Gaussian message $m_{f_{\boldsymbol{y}} \rightarrow \boldsymbol{X}}(\boldsymbol{X})$
with mean and covariance in \eqref{matrixXandSigmaX}.

\end{document}